\documentstyle[12pt,a4,epsf]{article} 
\newcommand{\be}{\begin{equation}}
\newcommand{\ee}{\end{equation}}
 
\newcommand{\bea}{\begin{eqnarray}}
\newcommand{\eea}{\end{eqnarray}}
\newcommand{\ba}{\begin{array}}
\newcommand{\ea}{\end{array}}
\newcommand{\lsim}{{\;\raise0.3ex\hbox{$<$\kern-0.75em
\raise-1.1ex\hbox{$\sim$}}
\;}}
\newcommand{\gsim}{{\;\raise0.3ex\hbox{$>$\kern-0.75em
\raise-1.1ex\hbox{$\sim$}}
\;}}

 \newcommand{\matr}{\left(
\begin{array}} \newcommand{\ematr}{\end{array} \right)}
\newcommand{\g}{\gamma}

\begin{document}

\noindent {\bf TESTING THE LEFT-RIGHT SYMMETRIC MODEL AT
LINEAR COLLIDER.}

\bigskip
 \noindent {\bf Katri Huitu}

\small\noindent Helsinki Institute of Physics,
\noindent University of Helsinki, Finland
 
\normalsize
\noindent {\bf Jukka Maalampi}

\small\noindent Department of Physics,
\noindent University of Helsinki, Finland

\normalsize
\noindent {\bf Aarre Pietil\"a}

\small\noindent Department of Applied Physics,
\noindent University of Turku, Finland

\normalsize
\noindent {\bf Martti Raidal}

\small\noindent Department of Theoretical Physics,
\noindent University of Valencia, Spain

\normalsize
\noindent {\bf Raimo Vuopionper\"a}

\small\noindent Helsinki Institute of  Physics,
\noindent University of Helsinki, Finland

\vspace{1.cm}

\noindent {\bf Abstract}

\small\noindent  We review possible tests of the left-right electroweak
model at future linear colliders, concentrating on signatures of the central
predictions of the model, i.e.  right-handed currents, massive neutrinos and
triplet Higgs bosons. We analyse processes in  
$e^+e^-,$ $e^-e^-$ and $e^-\gamma$
collision modes. We present the mass reaches for
the new particles at linear collider and sensitivities to their couplings.

\normalsize

\vspace{1.3cm}
\section*{The model}

The left-right symmetric model (LRM)
\cite{Pati} of electroweak interactions, based on  the
$SU(2)_L\times SU(2)_R\times U(1)_{B-L}\,$ symmetry, has
many predictions one can test at linear collider (LC). The
extended symmetry implies the existence of new gauge bosons,
$W^{\pm}_R$ and $Z^0_2$, which mediate V+A -type, i.e.
right-handed, weak interactions. It also requires more
complicated Higgs sector than in the Standard Model (SM) to
realize the two-step spontaneous breaking of the gauge symmetry down to
the unbroken $U(1)$ of QED. The breaking of
$SU(2)_L\times SU(2)_R\times U(1)_{B-L}\,$ to the SM
symmetry
$SU(2)_L\times U(1)_{Y}\,$ is arranged with an $SU(2)_R$
triplet scalar, the right-triplet 
$(\Delta^{++}_R, \Delta^+_R, \Delta^0_R)$.  The
breaking of the SM symmetry is due to a bidoublet scalar
multiplet consisting of a doublet and a conjugated doublet
of $SU(2)_L$. Also a left-triplet 
$(\Delta^{++}_L, \Delta^+_L, \Delta^0_L)$ may exist and 
contribute to this breaking, but the vev of $ \Delta^0_L$ 
is tightly bound by the $\rho$ parameter, or the mass ratio 
of the ordinary weak bosons. The fermion contents of the LRM is the same
as that of SM, except that right-handed neutrinos $N$
also exist. In the most natural version of the model
$N$'s are heavy as a result of the see-saw mechanism
\cite{seesaw}, with a mass comparable to the masses of
the new gauge bosons.

All the central predictions of the LRM, i.e. the new gauge
bosons
$W^{\pm}_R$ and $Z^0_2$, the Higgs triplet
$(\Delta^{++}_R, \Delta^+_R,
\Delta^0_R)$ and the right-handed neutrinos $N$, are
connected  intriguingly  to each other through the spontaneous breaking
of the LR symmetry. In the following we will consider the
production of $W_R$, $\Delta^{++}_R$, and $N$, which
will constitute the most effective probes of LRM at linear
collider.

\section*{Signals of $W^{\pm}_R$ }

According to the results of Tevatron, the mass of the new
charged gauge boson is constrained to be $M_{W_R}\gsim 650$
GeV \cite{CDFmass}. Although this bound can be
evaded for some choices of model parameters (a mass as low
as $\sim 300$ GeV being possible \cite{Lan}), it is probable that the
pair production of $W_R$'s in 
$e^+e^-$ \cite{MaPiVu}  and in $e^-e^-$ \cite{MaPiVu,rizzo} is 
kinematically excluded in a LC
with collision energy below 1 TeV.
 A single $W_R$ production via
$e^+e^-\to W_L^{\pm}W_R^{\mp}$ and $e^-e^-\to W_L^-W_R^-$ may
be kinematically viable but it is suppressed by the
smallness of the
$W_LW_R$ mixing \cite{Lan,MaPiVu}.

When the collision energy is sufficiently above 1 TeV, the pair production
of $W_R$'s via $e^+e^-\to W_R^{\pm}W_R^{\mp}$ and $e^-e^-\to W_R^-W_R^-$
may become possible. At $\sqrt{s}=1.6$ TeV the cross section for the 
 $W_R^{+}W_R^{-}$ production is at the level of 100 fb for $M_{W_R}=650 $ GeV
\cite{MaPiVu} and the mass reach will be practically up to the threshold
value of $\sqrt{s}/2$. The pair production of the same sign $W_R$'s in $e^-e^-$ 
collisions is an excellent place to probe the LR model. The reaction
is mediated by a Majorana neutrino in t-channel and the doubly charged
Higgs boson in s-channel. The right handed neutrino ($N$) and 
$W_R$ boson mass reach of LC due to this process is plotted in Fig. 1. It is assumed in
this  plot that the doubly charged Higgs is very heavy ($M_{\Delta^{--}}=5M_{N}$)
so that its contribution is insignificant in comparison with the contribution
of the neutrino. The direct 
$M_{W_R}$ reach  is up to $\sqrt{s}/2$ and the indirect $M_{N}$ reach up to
about 20 TeV.

In the electron-photon collision mode $W_R$ can be produced
via the reaction $e^-\gamma \to W_R^-N$, which is
kinematically accessible provided the right-handed neutrino $N$ is
light enough  \cite{HuMaRa}.  For
$\sqrt{s_{e\gamma}} = 730$ GeV (corresponding to
$\sqrt{s}=800$ GeV in the $e^+e^-$ mode), $M_{W_R}=650$ GeV
and $M_{N}\lsim 75$ GeV  the cross section is in the range 10 -- 20 fb
corresponding to some 50 -- 100 events for the luminosity of 50
fb$^{-1}$. The right-handed neutrino
 has most naturally a mass much larger than 75 GeV,
making the reaction kinematically forbidden. 
The limits on the  mixing of the left-handed neutrino $\nu$ with $N$
 still
allow 
 the kinematically more favourable reaction $e^-\gamma 
\to W_R^-\nu$ to have the cross section 
at observable level.

The production of $W_R$'s can be identified through its
decay to the ordinary $W_L$ boson and a neutrino.

\section*{Signals of the right-handed neutrino}

The heavy right-handed neutrinos can be produced in 
$e^+e^-\to \overline NN$ proceeding via  a $Z_2$
exchange in $s$-channel and a $W_R$ exchange in
$t$-channel \cite{MMV}. In LRM the heavy neutrino is 
a Majorana neutrino, for which the cross section of the pair
production is slightly smaller than for a Dirac neutrino, in
particular close to the threshold, due to the well-known
$\beta^3$ suppression. The LC with anticipated luminosities can probe
neutrino masses practically up to the kinematical limit
$\sqrt{s}/2$.  To probe the Dirac and Majorana nature of
neutrinos one can use the angular distributions. The process gives an indirect
probe of the mass $M_{W_R}$. For $\sqrt s=1.6$ TeV the probe is up to
about 4 TeV, as presented in Fig. 1.

A better mass reach, up to $\sqrt{s}$, for the 
right-handed neutrino is offered by the reaction 
$e^+e^-\to \overline N\nu$ whose cross section 
can be at a few fb level when the constraints on the 
neutrino mixing are taken into account \cite{Janusz}.

\section*{Signals of the triplet Higgs}

The left-triplet $\Delta_L$ and right-triplet  $\Delta_R$
Higgses transform as ({\bf 3},{\bf 1},2) and ({\bf 1},{\bf
3},2) under
$SU(2)_L\times SU(2)_R\times U(1)_{B-L}\,$, respectively.
The gauge symmetry  prevents these scalars from coupling to
quarks, and their couplings to fermions violate lepton
number by two units, $|\Delta L|=2$. For discovery the most
favourable are the doubly charged components
$\Delta_{L,R}^{++}$ due to their decay to an energetic
like-sign lepton pair, $\Delta_{L,R}^{++}\to
\ell^+\ell^+$.

In $e^+e^-$ collisions a single $\Delta_{L,R}^{++}$ is
produced via the reaction $e^+e^-\to
e^-l^-\Delta_{L,R}^{++}$ \cite{Lusignoli}. The cross
section depends on two unknown parameters, the mass of
$\Delta_{L,R}^{++}$ and  the strength $h_{\Delta}$ of the
lepton number violating
$\Delta_{L,R}\ell\ell$ coupling. Assuming
$h_{\Delta}=0.1$ and $M_{\Delta_{L,R}^{++}}\gsim 100$ GeV
the cross section is in the range $10 - 10^{3}$ fb.
 \cite{Barenboim}. About three
orders of magnitude more stringent limit is achievable for
the ratio
$|h_{\Delta}/M_{\Delta_{L,R}^{++}}|^2$ than the present
limit of $10^{-5}$ from the Bhabha scattering \cite{Bhabha}.

Another process where a single $\Delta_{L,R}^{++}$ is produced is 
$e^-\g  \rightarrow  l^+ \Delta^{--}$.
The primary lepton created in the process will remain undetected as it is
radiated almost parallel to  the beam axis. 
One cannot tell whether this particle is a positron, antimuon or antitau.  
Therefore,
the quantity which one can test in the reaction is actually the sum
$ h_{ee}^2+h_{e\mu}^2+h_{e\tau}^2$. 
Assuming the integrated luminosity of $e^-\g$ collisions to be
$L=$ 5, 10, 20, 40 fb$^{-1}$ for 
$\sqrt{s_{e\gamma}}=$ 330, 460, 730, 1450 GeV, respectively, 
and that for the discovery of $\Delta_R^{--}$ one needs
ten events, we obtain   the upper bounds plotted in Fig. 2. 
The sensitivity of LC will thus be
\be
h_{ee},\, h_{e\mu},\, h_{e\tau}\lsim 10^{-3}
\ee
for $M_{\Delta^{--}}\lsim \sqrt{s_{e\gamma}}$. 
Among the present constraints only
$h_{e\mu}h_{ee}<3.2\times 10^{-11}\;{\rm GeV}^{-2}\cdot M_{\Delta^{--}} ^2$ 
obtained from
the process $\mu\to\overline eee$ \cite{Bell},  can compete with these
bounds and only so 
at the low mass values. For the coupling $h_{e\tau}$ there does not
exist any bound from the present experiments.

The pair production of $\Delta_{L,R}^{++}$'s proceeds
through the s-channel exchange of the photon and the
standard and the heavy $Z$ bosons. The cross section scales
as $\beta^3/s$ as a function of the c.m. energy $\sqrt{s}$, where
$\beta$ is the velocity of the final state particles
\cite{Grifols,gunionDelta}. For
$\Delta_{L}^{++}\Delta_{L}^{--}$ production at
$\sqrt{s}=500$ GeV it is on the level of a few hundreds of
fb's up to the vicinity of the kinematical limit, for the
$\Delta_{R}^{++}\Delta_{R}^{--}$ production slightly less.

The kinematically most favoured decay modes of the doubly
charged scalars are those to like-sign lepton pairs, which
provide  excellent discovery signals.

\newpage

\noindent{\bf Figure caption}

{\bf Figure 1.} 
Sensitivity of the processes $e^-e^-\to W_R^-W_R^-$ and $e^+e^-\to
\overline NN$ on the masses of the heavy charged gauge boson $W_R$ and
the right-handed neutrino $N$ at the collision energy
$\sqrt{s}=1.6$ TeV assuming discovery limit of 0.1 fb and $e^+e^-$
luminosity of 200 fb$^{-1}$. 
The mass of the doubly charged triplet Higgs $\Delta_R$ is taken to
be $M_{\Delta^{--}} = 5M_N$. Dashed line is for Majorana neutrinos, solid line
for Dirac neutrinos.

{\bf Figure 2.} 
Sensitivity of the process $e^-\g  \rightarrow  l^+ \Delta^{--}$ 
on the lepton number
violating couplings $h_{ee}$, $h_{e\mu}$ and $h_{e\tau}$ 
as a function of the doubly
charged Higgs mass $M_{\Delta}$ for various collision energies.
The discovery limit of $\Delta^{--}$ is  assumed to be ten events.

\newpage

\begin{figure*}
\begin{center}
 \mbox{\epsfxsize=14cm\epsfysize=14cm\epsffile{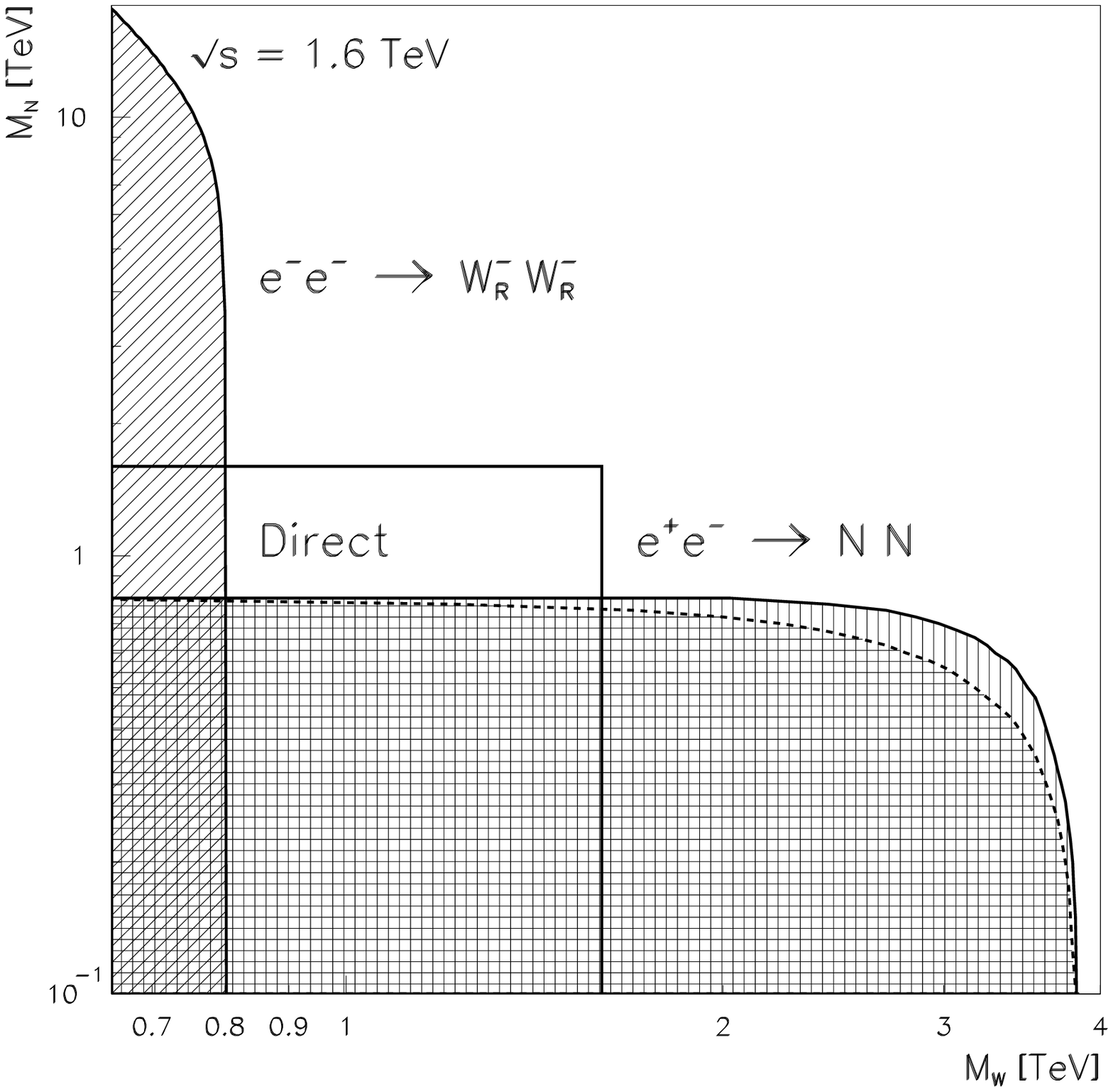}}
\caption{}
\end{center}
\end{figure*}

\begin{figure*}
\begin{center}
 \mbox{\epsfxsize=15cm\epsfysize=15cm\epsffile{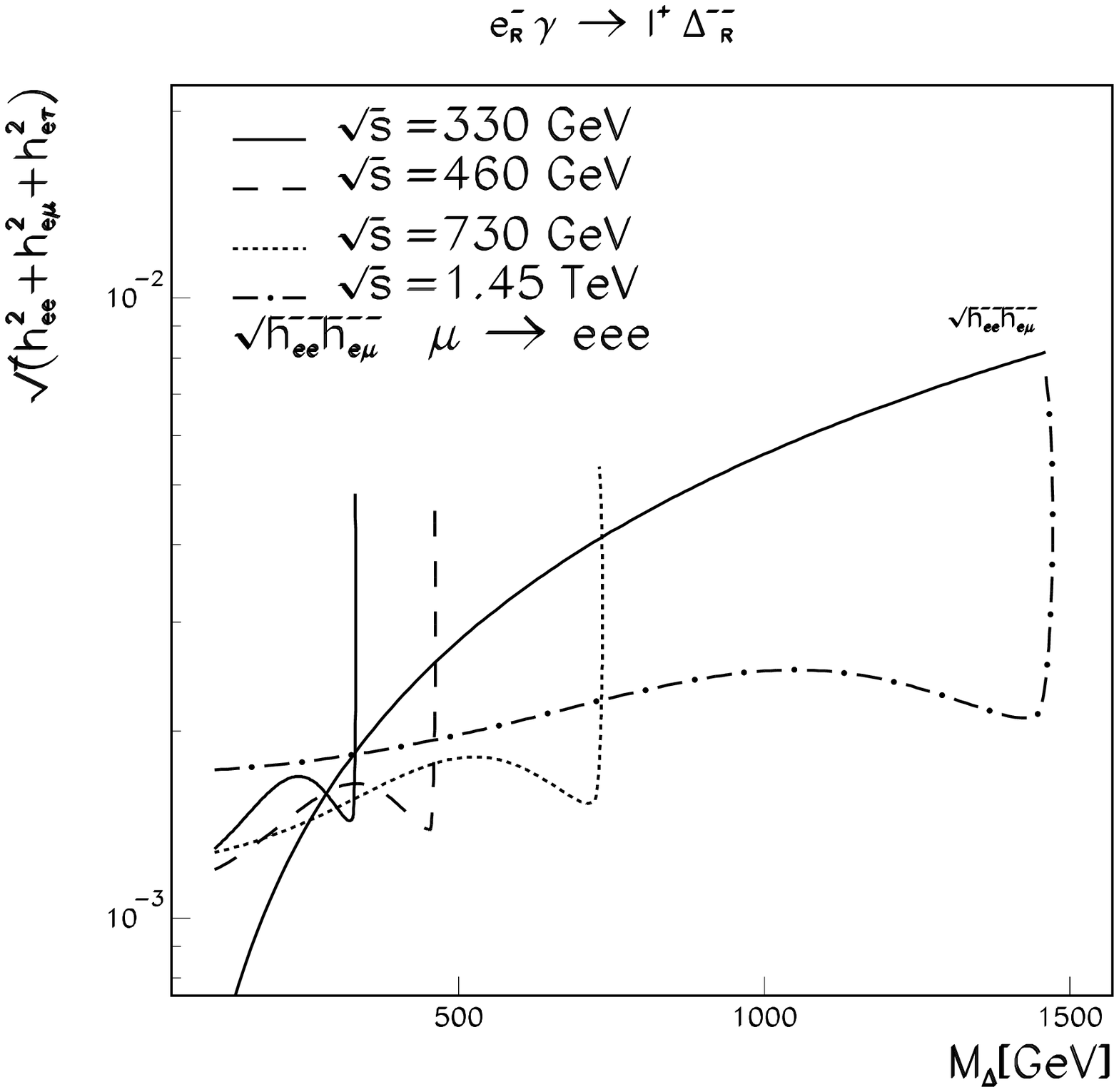}}
\caption{}
\end{center}
\end{figure*}

\end{document}